\documentclass[sigconf]{acmart}

\AtBeginDocument{%
  \providecommand\BibTeX{{%
    \normalfont B\kern-0.5em{\scshape i\kern-0.25em b}\kern-0.8em\TeX}}}

\setcopyright{acmcopyright}
\copyrightyear{2018}
\acmYear{2018}
\acmDOI{XXXXXXX.XXXXXXX}

\acmConference[Social Presence in Virtual Event Spaces (Workshop)]{CHI '22}{Apr 30--May 06,
  2022}{New Orleans, USA}
\acmPrice{15.00}
\acmISBN{978-1-4503-XXXX-X/18/06}

\begin{document}

%\title{Current and Emerging HCI Perspectives on Exploring Emotion and Technology}

%\title{Are HCI Perspectives Necessary for Research in Technology and Emotion?}
\title{Social Virtual Reality Avatar Biosignal Animations as Availability Status Indicators}

\author{Abdallah El Ali}
%\email{webmaster@marysville-ohio.com}
\affiliation{%
  \institution{Centrum Wiskunde \& Informatica (CWI)}
%  \streetaddress{P.O. Box 1212}
  \city{Amsterdam}
  \country{The Netherlands}
}
\email{aea@cwi.nl}

\author{Sueyoon Lee}
\affiliation{%
  \institution{Centrum Wiskunde \& Informatica (CWI)}
  \city{Amsterdam}
  \country{The Netherlands}}
\email{sueyoon@cwi.nl}

\author{Pablo Cesar}
\affiliation{%
  \institution{Centrum Wiskunde \& Informatica (CWI)}
  \institution{Delft University of Technology}
%   // Delft University of Technology}
%  \streetaddress{P.O. Box 1212}
  \city{Amsterdam}
  \country{The Netherlands}
}
\email{p.s.cesar@cwi.nl}

%\affiliation{%
%  \institution{Inria Paris-Rocquencourt}

\renewcommand{\shortauthors}{El Ali, et al.}

\begin{abstract}

In this position paper, we outline our research challenges in Affective Interactive Systems, and present recent work on visualizing avatar biosignals for social VR entertainment. We highlight considerations for how biosignals animations in social VR spaces can (falsely) indicate users' availability status.

%emotion Human-Computer Interaction (HCI), and zoom in on a case study of how developing real-time and continuous emotion self-report acquisition techniques necessitates HCI perspectives.

%some current and emerging research areas for exploring emotion and technology from an HCI perspective.

\end{abstract}


%\ccsdesc[300]{Virtual Reality}


\maketitle

\section{Our Research Challenges}
%A core aspect of such augmentation is the computational study of human emotion (affect), which are 

Within our Distributed \& Interactive Systems research group, we focus on designing and developing \textbf{Affective Interactive Systems}, where we draw on Virtual Reality (VR) systems as an experimental testbed to delve deeper into the links between technologically-mediated human affect and physiological signals. One specific research challenge we focus on is exploring \textit{Affective Augmentation} technology, where we ask: how can we develop systems that can augment our physical / virtual bodies and sensory perception to enhance our affective states and (social) interactions? As a step in this direction, we have recently started exploring biosignal sensing and visualization for avatars across social Virtual Reality (VR) spaces. In this position paper, we focus on visualizing avatar biosignals for social VR entertainment, and raise the question of how biosignals animations can (falsely) indicate users' availability status.

\section{Visualizing Avatar Biosignals for Social Virtual Reality Entertainment}

During real social interactions, we typically draw on a wide range of visible non-verbal behavioral cues (facial expressions, gestures, etc.) to form impressions and facilitate communication with one another \cite{Knapp2013}. However this remains a challenge for digitally mediated communication where several such cues are missing \cite{Walther2011,Kiesler1984}. Recent technological advances have shown that it is possible to reveal previously invisible physiological data ("biosignals") about others to better inform our impressions. These biosignals can provide useful insights to others about our internal emotional and cognitive states, where social sharing of such data allows others to peek into our normally hidden experiences. To this end, researchers have shown that "expressive" biosignals \cite{Liu2017}, where an individual's biosignals are displayed as a social cue, can allow us to better recognize others' and helps express our own emotional and physical state \cite{Liu2021,Min2014}, and can result in heightened co-presence \cite{Feijt2021}.

\begin{figure}
        \centering
        \includegraphics[width=1\linewidth]{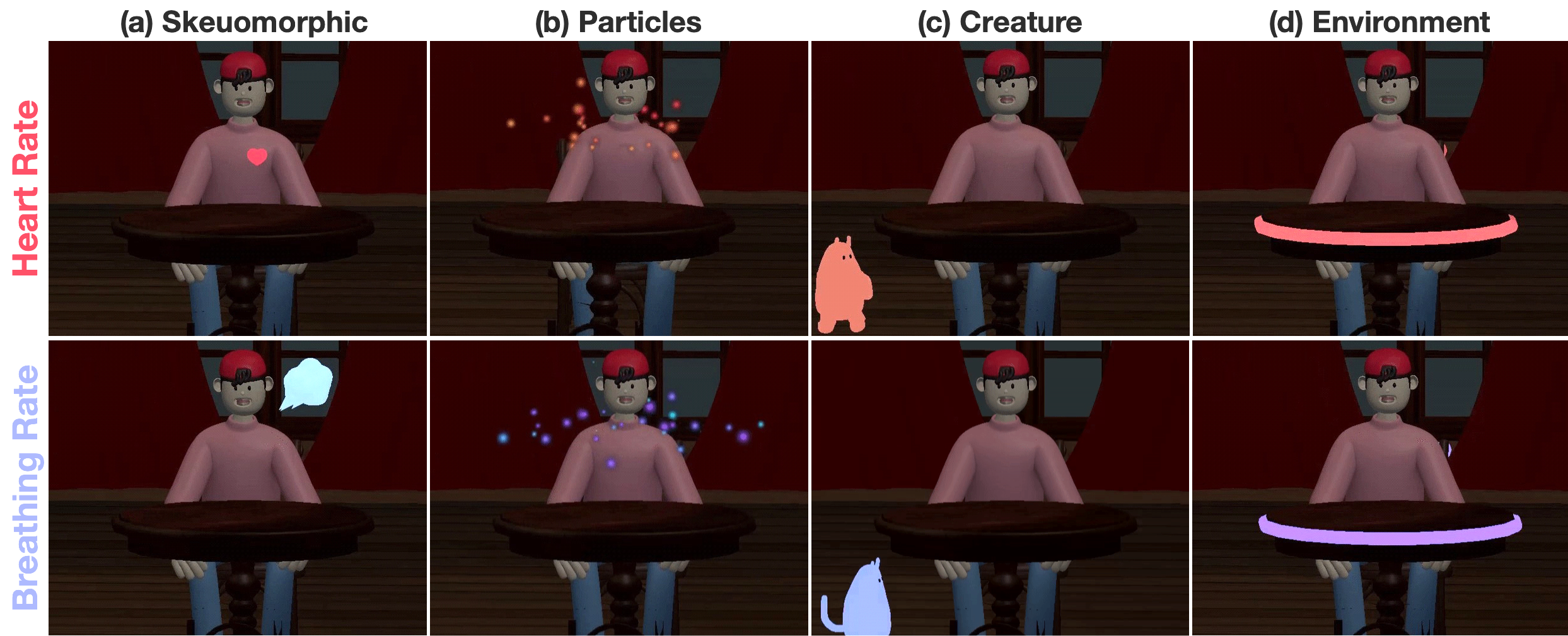}
                \caption{Unity-based visualizations for heart rate (top) and breathing rate (bottom).}                
        \label{fig:cover_img}
        \Description{Image showing our Unity-based visualizations for heart rate (top row) and breathing rate (bottom row).}
    \end{figure}
    
In our upcoming work \cite{Lee2022}, we explore how to design and visualize avatar biosignals in the context of social VR entertainment. Specifically, we asked: which are the most effective visualizations of Heart Rate (HR) and Breathing Rate (BR) biosignals in an immersive, virtual music event scenario? To answer this, we ran a controlled, within-subjects experiment (N=32) with pairs of users to investigate the effects of biosignals (Heart Rate vs. Breathing Rate), visualization (Skeuomorphic vs. Particles vs. Creature vs. Environment), and signal rate (Low vs. Rest vs. High) on perceived avatar arousal, user perceived distraction, and overall user attitudes and preferences. These visualizations can be seen in Fig. \ref{fig:cover_img}. We found that skeuomorphic visualizations for both biosignals allow differentiable arousal inference, skeuomorphic and particles were least distracting for HR visualization, whereas all were similarly distracting for BR, and that biosignal perceptions often depend on avatar relations, entertainment type, and emotion inference of avatars versus spaces. 

In our work, we found strong links between participants' responses towards avatar biosignals, and the role that visualizing biosignals on avatars (as animations) may play for increasing social presence. For this position paper, we focus on one aspect: availability status indicators. In our context mapping session, one theme that emerged was \textit{Presence of Self and Other}. Here, participants mentioned that within social VR spaces, it was common for them to prove their presence in VR via any means (e.g., sending emoji), as it was not always apparent to them if they were present (\textit{P5: "Let's just say someone's like daydreaming or like looking in the distance, it's kind of hard to see that in avatars."}). Similarly, participants complained about not being able to check the presence of others, where they had to actively engage with an avatar (\textit{P3: "...making sure that I'm making people aware that I'm there and able to talk."}). In our semi-structured interviews, one of the themes that emerged was \textit{Connecting with Others and Presence / Immersion}. Here, participants found biosignals as a means to better connect with others. For some, biosignal visualizations provided a feeling that who they spoke with is in fact a real person (\textit{P16: "I feel like I'm in a video game...but if you show a heartbeat or breathing...I feel like more connected."}). 

%    
%\begin{figure}[t]
%	\centering
%	\includegraphics[width=0.55\linewidth]{fig/research_approach.pdf}
%	\caption{Our two-part study approach. Contributions outlined in (\textbf{bold}) magenta.}
%	\Description{Flow diagram showing our study approach, which is split into two parts: Part 1 and Part 2. Gray boxes are the studies conducted, whereas the light orange boxes show the outcome of each study. Paper contributions are outlined in bold and magenta.}
%	\label{fig:approach}
%\end{figure}

\section{Animated Biosignals can (Falsely) Indicate Availability Status}

From our study, it became apparent that biosignals can play a role as avatar online status indicators (cf., awareness displays \cite{Dey2006}). This can provide another means for "idle" users to indicate their presence to others in a social virtual space, or even allow verifying humanness. However, this can also provide a new means to mislead others \cite{Cobb2020} in facilitating wasted interactions. Given this, one would need to consider freezing the biosignal when users are away (e.g., removing HMD), since the animations can provide the incorrect social cue at times. In this respect, the choice of how to visualize a biosignal and how it behaves through user interactions becomes paramount, where we caution that \textbf{biosignal animations can falsely indicate availability status.} While other non-verbal cues (e.g., fiddling motion, gaze) can also indicate online status, biosignals, by virtue of always being present, can falsely breathe life into an otherwise still avatar when the user is absent. 

We believe that as more users enter social VR spaces (the so-called "metaverse"), further consideration needs to be given on the role of biosignals for social presence, specifically on how to effectively display online availability statuses without misleading others. If this is done through communicating biosignals, whether expressive (e.g., heart rate creature animations \cite{Liu2021}) or through for example realistic blood flow animations \cite{McDuff2021}, it warrants further study on what happens to subsequent social interactions and emotional connections when the user leaves (whether by choice or not) their avatar, for shorter or longer periods.

\section{Author Biographies}

\textbf{Abdallah El Ali} is a research scientist at Centrum Wiskunde \& Informatica (CWI) in Amsterdam within the Distributed \& Interactive Systems (DIS) group. He is leading human-computer interaction (HCI) research within the Affective Interactive Systems research area. His focus is on ground truth label acquisition techniques, affective state visualization across environments (mobile, wearable, XR), and bio-responsive interactive prototypes. He is also part of the  executive board of CHI Nederland, the local SIGCHI chapter in the Netherlands. Website: \url{https://abdoelali.com}
\\

\noindent \textbf{Sueyoon Lee} is a research assistant at Centrum Wiskunde \& Informatica (CWI) in Amsterdam within the Distributed \& Interactive Systems (DIS) group. She follows a user-centric design approach for creating immersive yet comfortable experiences with design and technology. She currently leads design research within the TRACTION EU project, where her focus is on building a social VR lobby for post-watching of opera experiences. Website: \url{https://sueyoonlee.com}
\\

\noindent \textbf{Pablo Cesar} leads the Distributed and Interactive Systems Group, Centrum Wiskunde \& Informatica (CWI) and is Professor with TU Delft, The Netherlands. His research combines HCI and multimedia systems, and focuses on modelling and controlling complex collections of media objects (real-time media, sensor data) that are distributed in time and space. He has received the prestigious 2020 Netherlands Prize for ICT Research on human-centered multimedia systems, and is the principal investigator from CWI on topics of social virtual reality and affective computing. Website: \url{https://www.pablocesar.me}

%He is a member of the Editorial Board of the IEEE Multimedia, ACM Transactions on Multimedia, and IEEE Transactions of Multimedia, among others. He has acted as an Invited Expert at the European Commission's Future Media Internet Architecture Think Tank.

%\textbf{Abdallah El Ali} received his PhD degree from the University of Amsterdam in 2013. Currently, he is a tenure-track researcher at Centrum Wiskunde \& Informatica (CWI) in Amsterdam within the Distributed \& Interactive Systems (DIS) group. He is leading human-computer interaction (HCI) research within the Affective Interactive Systems research area. His focus is on ground truth label acquisition techniques, affective state visualization across environments (mobile, wearable, XR), and bio-responsive interactive prototypes. He is also part of the executive board of CHI Nederland, the local SIGCHI chapter in the Netherlands. Website: \url{https://abdoelali.com/}.

\bibliographystyle{ACM-Reference-Format}
\bibliography{socialpresence-ws-chi22}


\begin{thebibliography}{11}

%%% the Web address.

\ifx \showCODEN    \undefined \def \showCODEN     #1{\unskip}     \fi
\ifx \showDOI      \undefined \def \showDOI       #1{#1}\fi
\ifx \showISBNx    \undefined \def \showISBNx     #1{\unskip}     \fi
\ifx \showISBNxiii \undefined \def \showISBNxiii  #1{\unskip}     \fi
\ifx \showISSN     \undefined \def \showISSN      #1{\unskip}     \fi
\ifx \showLCCN     \undefined \def \showLCCN      #1{\unskip}     \fi
\ifx \shownote     \undefined \def \shownote      #1{#1}          \fi
\ifx \showarticletitle \undefined \def \showarticletitle #1{#1}   \fi
\ifx \showURL      \undefined \def \showURL       {\relax}        \fi
% invisible to TeX
\providecommand\bibfield[2]{#2}
\providecommand\bibinfo[2]{#2}
\providecommand\natexlab[1]{#1}
\providecommand\showeprint[2][]{arXiv:#2}

\bibitem[Cobb et~al\mbox{.}(2020)]%
        {Cobb2020}
\bibfield{author}{\bibinfo{person}{Camille Cobb}, \bibinfo{person}{Lucy Simko},
  \bibinfo{person}{Tadayoshi Kohno}, {and} \bibinfo{person}{Alexis Hiniker}.}
  \bibinfo{year}{2020}\natexlab{}.
\newblock \bibinfo{booktitle}{\emph{User Experiences with Online Status
  Indicators}}.
\newblock \bibinfo{publisher}{Association for Computing Machinery},
  \bibinfo{address}{New York, NY, USA}, \bibinfo{pages}{1?12}.
\newblock
\showISBNx{9781450367080}
\urldef\tempurl%
\url{https://doi.org/10.1145/3313831.3376240}
\showURL{%
\tempurl}


\bibitem[Dey and de~Guzman(2006)]%
        {Dey2006}
\bibfield{author}{\bibinfo{person}{Anind~K. Dey} {and} \bibinfo{person}{Ed de
  Guzman}.} \bibinfo{year}{2006}\natexlab{}.
\newblock \bibinfo{booktitle}{\emph{From Awareness to Connectedness: The Design
  and Deployment of Presence Displays}}.
\newblock \bibinfo{publisher}{Association for Computing Machinery},
  \bibinfo{address}{New York, NY, USA}, \bibinfo{pages}{899?908}.
\newblock
\showISBNx{1595933727}
\urldef\tempurl%
\url{https://doi.org/10.1145/1124772.1124905}
\showURL{%
\tempurl}


\bibitem[Feijt et~al\mbox{.}(2021)]%
        {Feijt2021}
\bibfield{author}{\bibinfo{person}{Milou~A. Feijt},
  \bibinfo{person}{Joyce~H.D.M. Westerink}, \bibinfo{person}{Yvonne A.W.~De
  Kort}, {and} \bibinfo{person}{Wijnand~A. IJsselsteijn}.}
  \bibinfo{year}{2021}\natexlab{}.
\newblock \showarticletitle{Sharing biosignals: An analysis of the experiential
  and communication properties of interpersonal psychophysiology}.
\newblock \bibinfo{journal}{\emph{Human–Computer Interaction}}
  \bibinfo{volume}{0}, \bibinfo{number}{0} (\bibinfo{year}{2021}),
  \bibinfo{pages}{1--30}.
\newblock
\urldef\tempurl%
\url{https://doi.org/10.1080/07370024.2021.1913164}
\showDOI{\tempurl}


\bibitem[Kiesler et~al\mbox{.}(1984)]%
        {Kiesler1984}
\bibfield{author}{\bibinfo{person}{S. Kiesler}, \bibinfo{person}{J. Siegel},
  {and} \bibinfo{person}{T. McGuire}.} \bibinfo{year}{1984}\natexlab{}.
\newblock \showarticletitle{Social psychological aspects of computer-mediated
  communication}.
\newblock \bibinfo{journal}{\emph{Computer Supported Cooperative Work}}
  (\bibinfo{year}{1984}), \bibinfo{pages}{657--682}.
\newblock


\bibitem[Knapp et~al\mbox{.}(2013)]%
        {Knapp2013}
\bibfield{author}{\bibinfo{person}{M.L. Knapp}, \bibinfo{person}{J.A. Hall},
  {and} \bibinfo{person}{T.G. Horgan}.} \bibinfo{year}{2013}\natexlab{}.
\newblock \bibinfo{booktitle}{\emph{Nonverbal Communication in Human
  Interaction}}.
\newblock \bibinfo{publisher}{Cengage Learning}.
\newblock
\showISBNx{9781133311591}
\showLCCN{2012946947}


\bibitem[Lee et~al\mbox{.}(2022)]%
        {Lee2022}
\bibfield{author}{\bibinfo{person}{Sueyoon Lee}, \bibinfo{person}{Abdallah
  El~Ali}, \bibinfo{person}{Maarten Wijntjes}, {and} \bibinfo{person}{Pablo
  Cesar}.} \bibinfo{year}{2022}\natexlab{}.
\newblock \showarticletitle{Understanding and Designing Avatar Biosignal
  Visualizations for Social Virtual Reality Entertainment}. In
  \bibinfo{booktitle}{\emph{Proc. CHI '22 (to be published)}} (New Orleans, LA,
  USA). \bibinfo{publisher}{ACM}, \bibinfo{address}{New York, NY, USA}.
\newblock
\showISBNx{978-1-4503-9157-3/22/04}
\urldef\tempurl%
\url{https://doi.org/10.1145/3491102.3517451}
\showURL{%
\tempurl}


\bibitem[Liu et~al\mbox{.}(2017)]%
        {Liu2017}
\bibfield{author}{\bibinfo{person}{Fannie Liu}, \bibinfo{person}{Laura
  Dabbish}, {and} \bibinfo{person}{Geoff Kaufman}.}
  \bibinfo{year}{2017}\natexlab{}.
\newblock \showarticletitle{Can Biosignals Be Expressive? How Visualizations
  Affect Impression Formation from Shared Brain Activity}.
\newblock \bibinfo{journal}{\emph{Proc. ACM Hum.-Comput. Interact.}}
  \bibinfo{volume}{1}, \bibinfo{number}{CSCW}, Article \bibinfo{articleno}{71}
  (\bibinfo{date}{Dec.} \bibinfo{year}{2017}), \bibinfo{numpages}{21}~pages.
\newblock
\urldef\tempurl%
\url{https://doi.org/10.1145/3134706}
\showDOI{\tempurl}


\bibitem[Liu et~al\mbox{.}(2021)]%
        {Liu2021}
\bibfield{author}{\bibinfo{person}{Fannie Liu}, \bibinfo{person}{Chunjong
  Park}, \bibinfo{person}{Yu~Jiang Tham}, \bibinfo{person}{Tsung-Yu Tsai},
  \bibinfo{person}{Laura Dabbish}, \bibinfo{person}{Geoff Kaufman}, {and}
  \bibinfo{person}{Andr\'{e}s Monroy-Hern\'{a}ndez}.}
  \bibinfo{year}{2021}\natexlab{}.
\newblock \showarticletitle{Significant Otter: Understanding the Role of
  Biosignals in Communication}. In \bibinfo{booktitle}{\emph{Proc. CHI '21}}
  (Yokohama, Japan). \bibinfo{publisher}{ACM}, \bibinfo{address}{NY, USA},
  Article \bibinfo{articleno}{334}, \bibinfo{numpages}{15}~pages.
\newblock
\showISBNx{9781450380966}
\urldef\tempurl%
\url{https://doi.org/10.1145/3411764.3445200}
\showDOI{\tempurl}


\bibitem[McDuff and Nowara(2021)]%
        {McDuff2021}
\bibfield{author}{\bibinfo{person}{Daniel~J. McDuff} {and}
  \bibinfo{person}{Ewa~Magdalena Nowara}.} \bibinfo{year}{2021}\natexlab{}.
\newblock \bibinfo{booktitle}{\emph{?Warm Bodies?: A Post-Processing Technique
  for Animating Dynamic Blood Flow on Photos and Avatars}}.
\newblock \bibinfo{publisher}{ACM}, \bibinfo{address}{New York, NY, USA}.
\newblock
\showISBNx{9781450380966}
\urldef\tempurl%
\url{https://doi.org/10.1145/3411764.3445719}
\showURL{%
\tempurl}


\bibitem[Min and Nam(2014)]%
        {Min2014}
\bibfield{author}{\bibinfo{person}{Hyeryung~Christine Min} {and}
  \bibinfo{person}{Tek-Jin Nam}.} \bibinfo{year}{2014}\natexlab{}.
\newblock \showarticletitle{Biosignal Sharing for Affective Connectedness}. In
  \bibinfo{booktitle}{\emph{CHI '14 Extended Abstracts on Human Factors in
  Computing Systems}} (Toronto, Ontario, Canada) \emph{(\bibinfo{series}{CHI EA
  '14})}. \bibinfo{publisher}{ACM}, \bibinfo{address}{New York, NY, USA},
  \bibinfo{pages}{2191?2196}.
\newblock
\showISBNx{9781450324748}
\urldef\tempurl%
\url{https://doi.org/10.1145/2559206.2581345}
\showDOI{\tempurl}


\bibitem[Walther(2011)]%
        {Walther2011}
\bibfield{author}{\bibinfo{person}{Joseph Walther}.}
  \bibinfo{year}{2011}\natexlab{}.
\newblock \showarticletitle{Theories of computer-mediated communication and
  interpersonal relations}.
\newblock \bibinfo{journal}{\emph{The Handbook of Interpersonal Communication}}
  (\bibinfo{date}{01} \bibinfo{year}{2011}), \bibinfo{pages}{443--479}.
\newblock


\end{thebibliography}


\end{document}